\documentclass[conference]{IEEEtran}
\IEEEoverridecommandlockouts

\usepackage{cite}
\usepackage{amsmath,amssymb,amsfonts}
\usepackage{graphicx}
\usepackage{tabularx}
\usepackage{array}
\usepackage{stfloats}
\graphicspath{ {./images/} }

\usepackage{textcomp}
\usepackage{multirow}
\usepackage{comment}
 \usepackage{booktabs}
\usepackage{flushend}
\usepackage{url}

\usepackage{algorithm2e}
\RestyleAlgo{ruled}

\usepackage[table,xcdraw]{xcolor}

\usepackage[a4paper, total={184mm,239mm}]{geometry}
\def\BibTeX{{\rm B\kern-.05em{\sc i\kern-.025em b}\kern-.08em
    T\kern-.1667em\lower.7ex\hbox{E}\kern-.125emX}}
\begin{document}

\title{Side-Channel Extraction of Dataflow AI Accelerator Hardware Parameters}

\author{
    \IEEEauthorblockN{
        Guillaume Lomet\IEEEauthorrefmark{1},
        Rubén Salvador\IEEEauthorrefmark{2},
        Brice Colombier\IEEEauthorrefmark{3},
        Vincent Grosso\IEEEauthorrefmark{3},
        Olivier Sentieys\IEEEauthorrefmark{1},
        Cédric Killian\IEEEauthorrefmark{3}
        }

     \IEEEauthorblockA{
        \IEEEauthorrefmark{1} Univ Rennes, Inria, IRISA, Rennes, France
    }
    \IEEEauthorblockA{
        \IEEEauthorrefmark{2} CentraleSupélec, Inria, CNRS, IRISA, Rennes, France
    }
    \IEEEauthorblockA{
        \IEEEauthorrefmark{3} Université Jean Monnet Saint-Etienne, CNRS, Institut d Optique Graduate School, Laboratoire Hubert Curien UMR 5516\\
        F-42023, SAINT-ETIENNE, France
    }
}

\maketitle

\begin{abstract}Dataflow neural network accelerators efficiently process AI tasks on FPGAs, with deployment simplified by ready-to-use frameworks and pre-trained models. However, this convenience makes them vulnerable to malicious actors seeking to reverse engineer valuable Intellectual Property (IP) through Side-Channel Attacks (SCA). This paper proposes a methodology to recover the hardware configuration of dataflow accelerators generated with the FINN framework. Through unsupervised dimensionality reduction, we reduce the computational overhead compared to the state-of-the-art, enabling lightweight classifiers to recover both folding and quantization parameters. We demonstrate an attack phase requiring only 337 ms to recover the hardware parameters with an accuracy of more than 95\% and 421 ms to fully recover these parameters with an averaging of 4 traces for a FINN-based accelerator running a CNN, both using a random forest classifier on side-channel traces, even with the accelerator dataflow fully loaded. This approach offers a more realistic attack scenario than existing methods, and compared to SoA attacks based on \textit{tsfresh}, our method requires 940$\times$ and 110$\times$ less time for preparation and attack phases, respectively, and gives better results even without averaging traces.

\end{abstract}
\vspace*{0.21cm}
\section{Introduction}
In recent years, artificial intelligence (AI) powered by deep neural networks (DNN) has become pervasive in various fields, including autonomous systems, image processing, the medical domain, or generative AI. However, these algorithms require substantial computational power and large datasets, driving the development of more powerful and efficient solutions beyond classical computer hardware (HW). One solution is using FPGAs, which can create reconfigurable accelerators optimized for specific tasks and requirements without the need to design a custom chip for each solution. 

In several use cases, such as video stream processing or large-scale image batch processing, dataflow accelerators have emerged as one of the most efficient solutions. FPGAs have also become more accessible through remote instances with shared HW and frameworks for end-to-end DNN design, reducing costs and providing ready-to-use development environments. However, DNN accelerators are valuable targets for attacks because of the high value of their intellectual property (IP) and the significant cost associated with creating and training models. Furthermore, remote FPGA instances like cloud servers or embedded systems are vulnerable to physical attacks, which can bypass traditional security measures within these devices~\cite{stojilovicVisionaryLookSecurity2023}.

Attacking DNNs using side-channel attacks (SCA) follows a sequence of steps, where each stage requires specific knowledge depending on its position in the attack flow. To penetrate the black box of a DNN accelerator, the first crucial element to understand is the HW architecture, which is the root of all the SCA strategies, from model recovery to stealing inputs. In FPGAs, this architecture is configurable, and users can tune parameters such as the parallelism or the arithmetic used.
Although these parameters need to be captured to mount a successful weight recovery through SCAs, most attack models assume the attacker knows the HW architecture.

This work \textbf{proposes a methodology to remotely reverse-engineer elements of the HW architecture in a dataflow accelerator implemented on FPGAs.} This is achieved by comparing the traces captured from the accelerator to a dataset of patterns corresponding to traces from accelerators with known architectures. The attack uses unsupervised dimensionality reduction of the traces, and then the transformed data are classified by each HW parameter. Our contributions include: 
\begin{itemize}
    \item To the best of our knowledge, we are the first to \textbf{recover multiple HW parameters simultaneously from SCA traces}: the folding, which is the parallelization of data (SIMD) and processing elements (PE), and the quantization used in the accelerator. For this study, we only focus on the \textbf{folding of the PE and the quantization}, and with an averaging of 4 traces with random inputs to reduce noise, we can fully recover these parameters.
    \item Using an \textbf{unsupervised dimensionality reduction} technique like principal component analysis (PCA), \textbf{we lower the computational cost}, enabling the use of more classifiers for extracting the HW parameters and improving the performance of the parameter recovery. 
    \item Compared to the state-of-the-art (SoA), \textbf{our methodology enables a more realistic attack model} and can be used as a \textbf{first step to learning the HW architecture}, which was an assumption made by previous attacks.
\end{itemize}

This paper is organized as follows: Section~\ref{background} discusses the necessary background,
including SCA on FPGA NN accelerators. Section~\ref{attack_model} presents the attacker model,
while Section~\ref{methodology} explains the experimental methodology. Results are presented in Section~\ref{experiments} before concluding in Section~\ref{conclusion}. 

\section{Background} \label{background}

\subsection{DNN accelerators on FPGA}
Two framework types are typically used to accelerate DNN computation on FPGA. First, sequential accelerators compute specific operations like tensor multiplications and activations layer-by-layer, similar to CPUs/GPUs (e.g., Vitis AI, TVM-VTA~\cite{moreau_hardware-software_2018}, Gemmini~\cite{gemmini-dac}). Second, dataflow accelerators process layers in parallel as soon as inputs are available (e.g., FINN~\cite{umuroglu_finn_2016}).
Dataflow accelerators have proven efficient for tasks like large batch or data stream inference~\cite{venieris_toolflows_2018}, often outperforming GPUs or specialized ASICs~\cite{hamanaka_exploration_2023}. However, they require more custom work during DNN model creation, increasing their IP value and thus interest to be recovered through, e.g., SCA.

\subsection{Remote sensors in FPGA for SCA}\label{soa_remote}

To deploy FPGA DNN accelerators, remote solutions like in datacenters~\cite{amazon_f1} or embedded systems optimize FPGA usage by sharing resources, reducing costs, and standardizing tools. Remote SCAs are possible via embedded sensors~\cite{glamocanin_are_2020}, simulating power analysis. While tools protect against some sensors like ring oscillators, others using Time-to-Digital Converters (TDC)~\cite{schellenbergInsideJob2018} or arithmetic unit misuse~\cite{meyers_stealthy_2023} remain hard to mitigate.
Though these sensors are challenging to calibrate remotely, recent studies implement automatic calibration systems, helping attackers get the right sensitivity using TDCs~\cite{udugama_VITI_TDCSelfCalibrating_2022,gravellier:hal-02380092}.

\subsection{SCA for DNN reverse engineering on FPGA}

Valuable data and IP, like the architecture and trained weights from DNN models, can be extracted via SCA~\cite{batina_csi_2019}. Sensitive inputs, e.g., medical data, can also be recovered using knowledge of the trained model~\cite{maji_leaky_2021}. However, FPGA’s reconfigurable HW makes SCAs strongly dependent on knowing the accelerator’s architecture. We identify four categories of data and IP recovery attacks through SCA on FPGA, ranging from closest to black-box to closest to white-box scenarios.

\subsubsection{Hardware architecture}
Attacks on the HW architecture include the framework type used~\cite{tianRemotePowerAttackVTA2021,gongye_SCARevEngIP_2023}, data and processing parallelism~\cite{meyersRevEngNNFolding2022}, and the arithmetic type used (number representation, precision). To the best of our knowledge, there is still no work in the last category, which is one of the focuses of this work. Unlike other categories, attacks on the HW architecture are based on reverse engineering using SCA (SCARE)~\cite{hutchison_improved_2007} on the circuit, rather than retrieving secrets.

\subsubsection{Model architecture}
These attacks target elements of the neural network (NN) model: number of layers, type of layers, model structure, and layer configuration~\cite{tianRemotePowerAttackVTA2021, yli-mayry_ExtractionBNN_EMSCA_2021}. 

\subsubsection{Model parameters} Several works focus on SCA to extract model parameters, including weights and normalization~\cite{gongye_SCARevEngIP_2023, yli-mayry_ExtractionBNN_EMSCA_2021, weerasena_revealing_2023, yoshidaModelReverseEngineering2021}. 
\subsubsection{Inputs} The last category considers retrieving inputs used at inference time~\cite{weiKnowWhatYouSee2018}~\cite{moiniRemoteSCABNN2021}.

To get closer to a black-box scenario, attackers need to know the HW architecture of the accelerator, as parallelism can disturb model recovery due to increased noise~\cite{meyersRevEngNNFolding2022}. Some attacks are linked to the arithmetic used~\cite{maji_leaky_2021}, or exploit information from the implementation tool~\cite{gongye_SCARevEngIP_2023}. 
The closest work~\cite{meyers_stealthy_2023} to our approach to extracting HW parameters from dataflow accelerators also targets FINN~\cite{umuroglu_finn_2016}. It extracts the number of parallel PEs in a matrix-vector-accumulation unit to aid in recovering neurons in fully connected layers, using a profiling attack based on power traces and a machine learning classifier. Despite their key role in model parameter extraction, arithmetic parameters have yet to be known during the SCA. This attack \cite{meyers_stealthy_2023} also requires computationally heavy statistical analysis, which will be difficult to generalize to other HW parameters.
Similarly, EM-based SCA reverse-engineering was used to recover the structure and operation scheduling (input partitioning, channel/pixel-level parallelism, and spatial mapping of input/output channels and HW) of the AMD-Xilinx DPU encrypted DNN commercial accelerator~\cite{gongye_SCARevEngIP_2023}.

\emph{Conversely, our attack is the first to recover the arithmetic type (number representation, precision) [cat. 1] and to relax the assumption of knowing the HW architecture [cat. 2, 3].}

\subsection{Machine Learning based SCA}

Machine learning has been shown to improve  SCA attacks, e.g.,  using convolutional neural networks (CNN) for profiling attacks without trace realignment~\cite{fischer_convolutional_2017}. 
Another approach reduces trace dimensions by transforming them into smaller vectors for profiling. Early works on cryptographic accelerators have used RNN-based auto-encoders~\cite{ramezanpour_scaul_2020, liu_side-channel_2022} to reduce trace dimensionality, improving noise elimination by preserving key features.
However, to recover DNN elements in FPGA accelerators, the primary method used is feature selection~\cite{zhang_stealing_2021, meyersRevEngNNFolding2022} using \textit{tsfresh}~\cite{christ_time_2018}. While this method is effective and relies on deterministic algorithms, (1) its reliability depends on fixed statistical elements, which can sometimes limit the number of relevant features~\cite{henderson_empirical_2021}; and (2) it takes up to three orders of magnitude more time compared to our approach using PCA and lightweight classifiers, as we show in this work.

\section{Attacker Model} \label{attack_model}
We focus on a DNN accelerator embedded in an FPGA with remote sensors, aiming to recover key parameters such as the folding strategy and quantization levels used in the victim's model through SCA. Table~\ref{tab:attack_model} summarizes the attacker's knowledge, capabilities, constraints, and results.

The attacker is assumed to have knowledge of the framework used, as well as all the configurable accelerator parameters. Additionally, the attacker can deploy a clock-synchronized sensor on the same FPGA, allowing precise monitoring of the accelerator's activity through power side channels. This synchronized setup provides the attacker with a detailed view of the accelerator’s internal operations.

Our method stands out from SoA approaches \cite{meyersRevEngNNFolding2022} in several key ways, offering a model that requires less effort from the attacker, making it easier to implement and thus more effective.
In our case, the data is processed in batches, meaning that when we capture power traces, the accelerator's dataflow pipeline is already fully loaded. 
In this scenario, the noise and complexity increase, making it harder to interpret traces directly. 
Additionally, since we do not have access to the input data, we cannot capture repeated traces of the same data for averaging but only traces with different inputs; however, we show how this still helps to reduce noise.
Lastly, the CPU in the SoC FPGA remains active during our attack running Linux OS and a Jupyter server, 
introducing additional noise into the traces. This makes our approach more realistic, as it better reflects typical deployments on remote FPGAs where the CPU is not isolated, leading to non ideal and noisier conditions.

Despite these challenges, our method remains effective. We successfully recover the folding and quantization parameters by capturing traces in this noisier, fully loaded dataflow state and using a data-independent approach. This highlights the robustness of our attack in extracting several key model details, even in more complex and realistic FPGA environments.

\begin{table}[!t]
    \centering
    \caption{Comparison of attacker models in \cite{meyersRevEngNNFolding2022} and this work.}
    \scriptsize
    \setlength{\tabcolsep}{5pt}
    \begin{tabularx}{\columnwidth}{|c|>{\arraybackslash}X|>{\centering\arraybackslash}m{6em}|>{\centering\arraybackslash}m{4em}|}
        \hline
        \multicolumn{2}{|c|}{\textbf{Attack Parameter}} & \textbf{Our method} & \textbf{\cite{meyersRevEngNNFolding2022}} \\
        \hline
        \multirow{3}{*}{\shortstack{Attacker\\knowledge}}& Framework used
        & Yes & Yes \\
        \cline{2-4}
        & Possible configurations of the victim
        & Yes & Yes \\
        \cline{2-4}
        & Victim model
        & Yes & Yes \\
        \hline
        \multirow{2}{*}{Constraints}& \textbf{CPU activity impact traces}
        & \textbf{Yes} & \textbf{No} \\
        \cline{2-4}
        & \textbf{Capture traces in loaded dataflow}
        & \textbf{Yes} & \textbf{No} \\
        \hline
        \multirow{3}{*}{\shortstack{Attacker\\capabilities}}& Can implement synchronized sensors
        & Yes & Yes \\
        \cline{2-4}
        & Attacker has access to inputs/outputs
        & No & No \\
        \cline{2-4}
        & \textbf{Uses batches of inputs}
        & \textbf{Yes} & \textbf{No} \\
        \hline
        \multirow{2}{*}{Results}& Recovers the folding
        & Yes & Yes \\
        \cline{2-4}
        & \textbf{Recovers the quantization}
        & \textbf{Yes} & \textbf{No} \\
        \hline
    \end{tabularx}
    \label{tab:attack_model}
 \vspace{-0.2cm}
\end{table}

\begin{figure}[!t]
    \centering
    \includegraphics[width=\columnwidth]{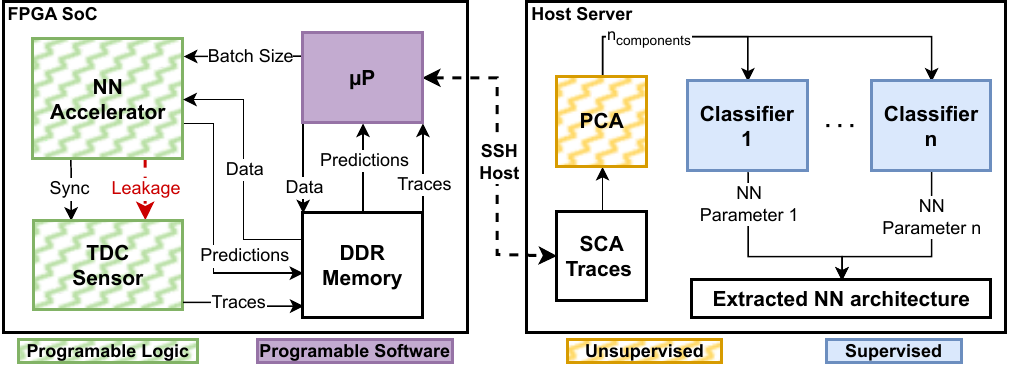}
    \vspace{-0.3cm}
    \caption{Overview of the attack system.}
    \label{fig:analyzer_layout}
    \vspace{-0.5cm}
\end{figure}

\section{Methodology} \label{methodology}

Our proposed attack methodology targets an implementation of a DNN accelerated on an FPGA, as illustrated in~Fig.\ref{fig:analyzer_layout}, and requires the use of a remote sensor located on the same chip as the accelerator and synchronized with the latter. Through leakage, the sensor will capture the power activity resulting from the calculations of the DNN accelerator. 

The proposed method consists of three phases as outlined in Algorithm~\ref{alg:attack_phases}: i) trace acquisition, ii) offline attack preparation, and iii) online attack. Traces captured during trace acquisition by a remote sensor are stored in the FPGA and then transferred via SSH for analysis. The attack involves dimensionality reduction of the traces via PCA and classification to identify the accelerator parameters, with one classifier used for each parameter type. The method is detailed in the following.

\subsection{Trace collection}\label{methodo:traces} 

The acquisition phase consists of two stages: first, creating the labeled database DB for attack preparation, which requires the ability to choose the accelerator parameters; second, performing the attack to recover these parameters without prior knowledge of the configuration.
In both phases, the acquired traces are preprocessed as follows:
\subsubsection{Windowing traces} Traces must contain at least all the samples corresponding to the the slowest stage of the dataflow DNN accelerator. This stage timing can be determined, for instance, through the FINN accelerator generation reports. During dataset construction, we set the minimum number of samples $N_{\text{s\_min}}$ to capture all relevant data from the slowest accelerator considered in our dataset, i.e., the one with less parallelism. This is computed using the slowest stage latency $T_{\text{dataflow}}$ and the sensor frequency $F_{\text{s}}$ from $N_{\text{s\_min}} = T_{\text{dataflow}} \times F_{\text{s}}$.

\subsubsection{Removing dataflow loading}\label{methodo:dataflow} Initial samples are skipped to ensure the dataflow is fully loaded. In a remote context, we consider that when triggering a blind measurement, it is unlikely that this would start at the exact same time as the accelerator with an empty dataflow. During the dataflow's loading phase, there is less noise, which would facilitate SCAs. Thus, in our case, our attack remains effective even in a more complex scenario where each stage of the dataflow is active simultaneously and loaded with different data. This is computed using the latency of one complete inference $T_{\text{inf}}$ of the slowest accelerator with $N_{\text{load}} = F_{\text{s}}\times T_{\text{inf}}$.
The two previous steps are referred to as \textit{Trim} the trace in Algorithm~\ref{alg:attack_phases}.

\subsubsection{Averaging traces} Averaging traces by a factor $n_{\text{average}}$ helps reduce noise, such as that from the CPU on SoC FPGAs. Since our attack model does not allow for input control, averaging is done between traces with different random inputs from the same victim. This parameter must be determined during the offline phase, as explained below.

\subsubsection{Normalizing traces} To remove the artifacts due to the automatic sensor calibration, we remove the mean of each trace:
\begin{gather*}
    x_{\text{normalized}} = x_i - \overline{X}  , \quad \forall i \in \{1, 2, \dots, n\}
\end{gather*}
where \( x_{\text{normalized}} \) represents the normalized value of the trace, \( x_i \) is the \(i\)-th time sample in the trace and \( \overline{X} \) is the mean of all samples in the trace

\subsection{Features extraction and classification}\label{methodo:attack}

The core idea of our attack methodology consists of two parts, as detailed in Algorithm~\ref{alg:attack_phases}: i) Dimensionality reduction of traces using feature extraction with PCA, ii) Classification of the extracted features using the dataset labels.

The PCA transformation is an unsupervised process that reduces the dimensionality of the traces while amplifying the variability between traces corresponding to different DNN parameter configurations. 
PCA computation, which transforms the trace samples into $n_\text{comp}$ (number of components), requires a fitting step performed during the preparation phase. It is described next.

\subsubsection{Offline phase (preparation)}

The PCA is computed \textit{(fit)} on a part of the database DB$_{\text{training}}$ and produces \textit{(transform)} a dimension reduction from $N_{\text{s\_min}}$ samples to $n_\text{comp}$ PCA components, which must be selected according to the classifier performance.
Regarding the selected classifier type, we tune it through a grid search to find the best combination of $n_\text{comp}$ and classifier parameters. For each combination, we measure the accuracy of the prediction thanks to DB. 
Finally, we perform a validation of the best combination as follows using a PCA reduction (transform) on DB$_{\text{test}}$ followed by a classification. We keep the combination of the results with the best accuracy.

\subsubsection{Online phase (attack)}
To perform the attack, we acquire one or several traces, followed by preprocessing and PCA transformation. The $n_\text{comp}$ are then passed to the classifier, which outputs the most probable accelerator configuration.

\begin{algorithm}
    \SetAlgoLined
    \caption{Acquisition and Attack Phases}
    \label{alg:attack_phases}
    \footnotesize
    \KwOut{DB$_{\text{training}}$ and DB$_{\text{test}}$}
    \SetKwBlock{Begin}{Acquisitions}{}
    \SetAlgoNoLine
    \Begin{
        \SetAlgoVlined
        \For{all possible NN accelerator configurations}{
            \Repeat{$N_{\text{traces}}$}{
                capture a side-channel trace\;
                label the trace\;
            }
        }
        Split DB into DB$_{\text{training}}$ and DB$_{\text{test}}$\;
    }
    
    \SetKwBlock{Begin}{Offline phase (preparation)}{}
    \Begin{
        \KwIn{DB$_{\text{training}}$, DB$_{\text{test}}$, $n_{\text{comp\_max}}$, $n_{\text{average}}$}
        \KwOut{Configuration with the best test accuracy}
        
        \SetKwBlock{Begin}{Common preprocessing (trimming + averaging + PCA)}{}
        \Begin{
            \SetAlgoVlined
            Trim the trace to the Normalized Window Size\;
            Normalize traces\;
            \If{averaging is required}{
                average over $n_{\text{average}}$ traces\;
            }
            Apply PCA on DB$_{\text{training}}$ and keep the top $n_{\text{comp\_max}}$ components\;
        }
        
        \SetKwBlock{Begin}{Classifier tuning}{}
        \Begin{
            \SetAlgoVlined
            \For{$n_{\text{comp}}\in[1, n_{\text{comp\_max}}]$}{
                Keep $n_{\text{comp}}$ components\;
                \For{selected classifier configurations (grid search)}{
                    Tune the classifier on DB$_{\text{training}}$ with the selected configuration\;
                    Save $(n_{\text{comp}}, \text{config\_classifier}, \text{accuracy})$\;
                }
            }
        }

        \SetKwBlock{Begin}{Classifier validation}{}
        \Begin{
            \SetAlgoVlined
            Apply PCA on DB$_{\text{test}}$\;
            \For{classifier configurations with best accuracy}{
                Classify DB$_{\text{test}}$\;
            }
        }
    }
        
    \SetKwBlock{Begin}{Online phase (attack)}{}
    \Begin{
        \SetAlgoVlined
        \KwIn{config\_classifier, $n_{\text{comp}}$, $n_{\text{average}}$}
        \KwOut{Most probable NN accelerator configuration}
        
        \For{each parameter of the NN accelerator}{
            Capture a new side-channel trace (or more)\;
            Apply preprocessing with $n_{\text{comp}}$ and $n_{\text{average}}$\;
            Perform classification with config\_classifier\;
        }
    }
\end{algorithm}

\section{Experimental Results} \label{experiments}

\subsection{Experimental Setup} \label{exp:setup}
\begin{figure}
 \vspace{-0.2cm}
    \centering
    \includegraphics[width=0.9\columnwidth]{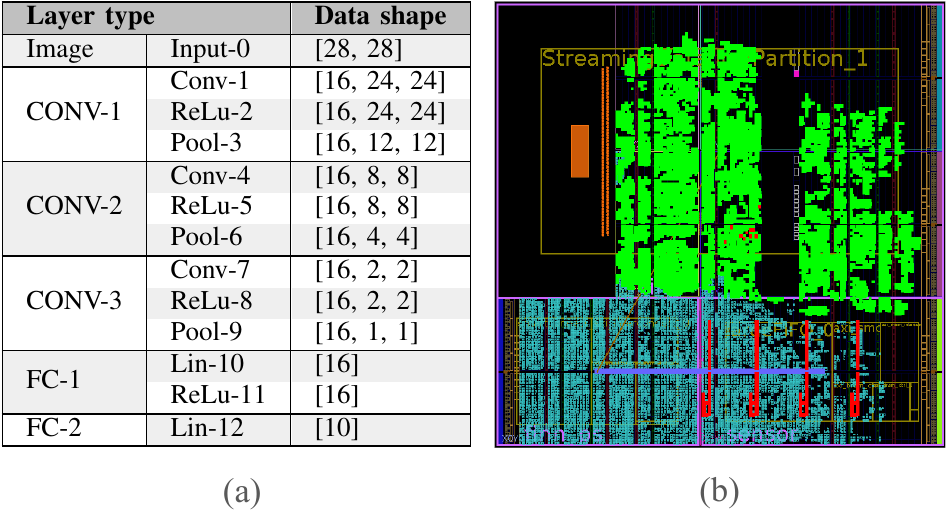}
    \vspace{-0.2cm}
    \caption{Overview of the experimental setup. (a) CNN victim model, (b) FPGA implementation of FINN victim accelerator:  accelerator in green, TDC sensors in red, communication components in light blue.    }
    \label{fig:experimental-setup}
    \vspace{-0.5cm}
\end{figure}

We used a PYNQ-Z2 FPGA board with a Xilinx Z7020 SoC containing an Artix-7 FPGA and a dual-core Arm Cortex-A9 CPU, both sharing 512 MB of DDR3 RAM and running Ubuntu OS with a Jupyter Notebook server. 
The considered DNN model is a 5-layer CNN trained on the MNIST dataset (28x28-pixel monochromatic images) to recognize handwritten digits, as detailed in Fig.~\ref{fig:experimental-setup}(a). The associated victim accelerator is based on FINN~(version 0.9) with the following HW configuration: no DSP, same folding and quantization size for all layers, and all parameters stored on embedded cache~(\mbox{LUTRAM} and BRAM) at synthesis. Only the Direct Memory Access~(DMA) used to push data in the input and to pull the data from the output of the accelerators communicates with the DRAM and the CPU. The designed accelerator parameters are: folding {1x, 2x, 4x, 8x} and quantization {4-bits, 6-bits} for a total of 8 different accelerator versions.

For the sensors, we use a modified version of the TDC from SCABox~\cite{gravellier:hal-02380092} with a length of 128 latches. Both the accelerator and TDC sensors use a clock frequency of 100 MHz. To capture the TDC sensor output, we added a DMA module to store the captured traces in the external DRAM memory. The DMA ensures synchronized capture of traces with the accelerator when it starts. 
\figurename~\ref{fig:experimental-setup}(b) shows the complete system floorplan.

\subsection{Data Acquisition}

Following the methodology from Section~\ref{methodology}, through trace observations and FINN reports, we estimate the minimal length for the sample window around 129k samples. Also, as explained in Section \ref{methodo:traces}, we trim the dataflow loading phase from the trace, corresponding to $129\text{k}\times2=258\text{k}$ samples. 

All the traces are collected on the FPGA and then sent through SSH after each acquisition. For our experiments, $N_{\text{traces}}=800$ traces are collected per accelerator configuration to have enough data for our dataset. The final dataset of traces includes a total of $N_\text{total}$ traces with
\begin{gather*}
    N_\text{total} = N_{\text{traces}} \times N_{\text{folding}} \times N_{\text{quantization}} 
\end{gather*}
where $N_{\text{folding}}$ and $N_{\text{quantization}}$ are the number of possible folding and quantization, respectively.
In our case study $N_{\text{folding}} = 4$,  $N_{\text{quantization}} = 2$, hence a total of 6400 traces for DB. We use 5120 traces for DB$_{\text{training}}$  and 1280 traces DB$_{\text{test}}$, balanced within all the configurations. 

\begin{figure}
    \centering
    \includegraphics[width=0.9\columnwidth]
    {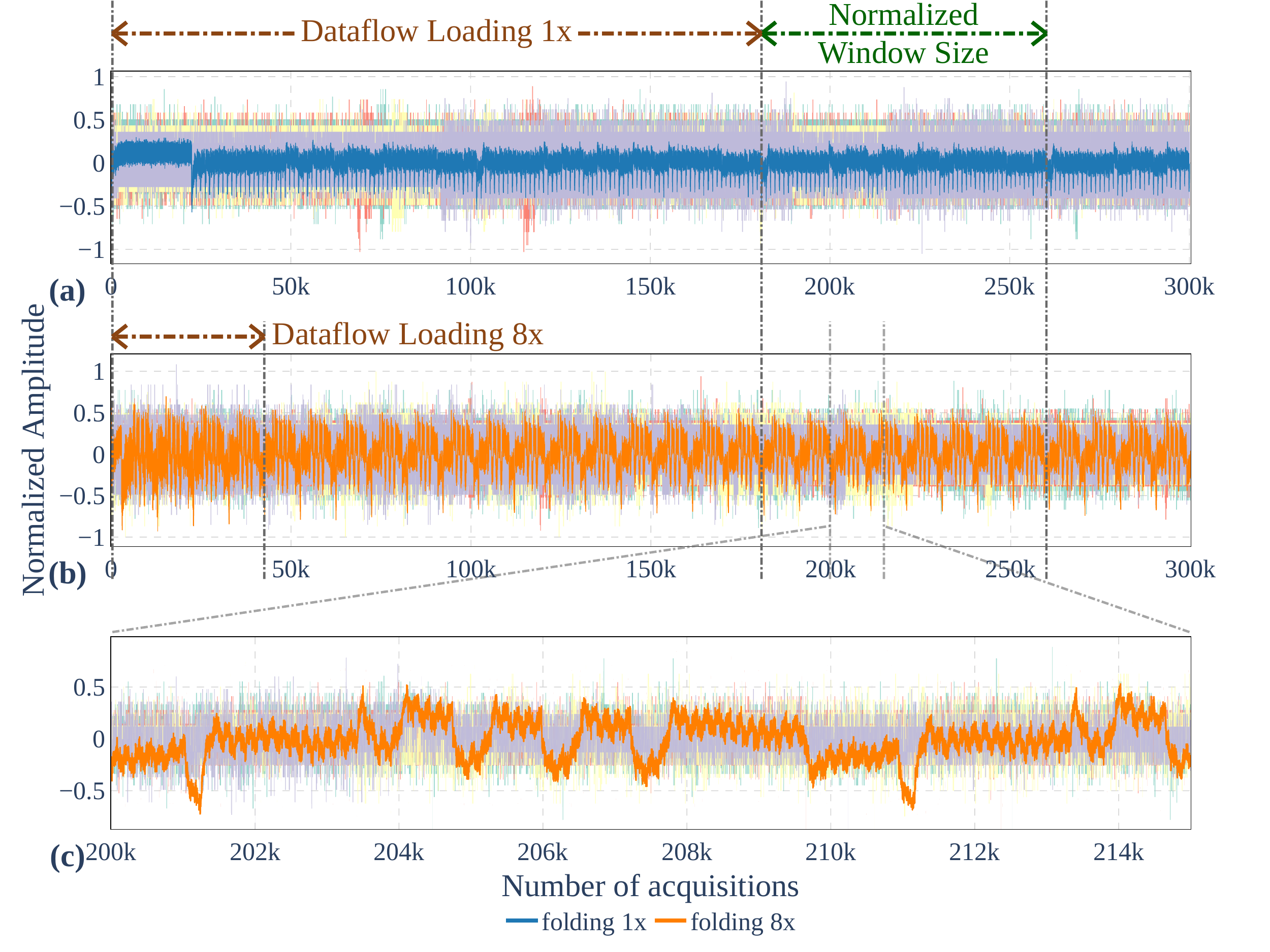}
    \vspace{-0.2cm}
    \caption{Average of 800 traces for 6-bit quantization and folding 1$\times$~(a) and 8$\times$~(b) and (c). Light-coloured traces represent raw traces without averaging.}
    \label{fig:traces}
    \vspace{-0.4cm}
\end{figure}
Fig.~\ref{fig:traces} shows an example trace with 6-bit quantization and folding factors of 1 and 8. To improve visibility in the figure, we reduced noise by averaging 800 traces. However, in the attack phase, averaging is limited to 1 to 4 traces in our case study. Without this averaging step the difference between the cases is not visible at first sight.
These traces reveal a repeating pattern, with the pattern frequency proportional to the folding factor. This clearly indicates that the DNN accelerator's power consumption leaks information through side channels.

Trace capture is triggered when the accelerator is launched, so the initial samples capture the NN's activity with an empty dataflow. This loading phase, labelled as \emph{Dataflow Loading} in Fig.~\ref{fig:traces}, has a duration inversely proportional to the folding factor and corresponds to the time taken for a single inference on the same data. 
As detailed in Section~\ref{methodology}, remote SCAs are unlikely to capture this event, so the dataflow loading regions are excluded from the acquired samples in the experiments. 

Fig.~\ref{fig:traces} highlights the \emph{Normalized Window Size}, corresponding to the duration of the slowest stage of the dataflow in the slowest accelerator configuration (here 1$\times$ folding, no parallelism). During this period, other dataflow stages continue operating with different data, allowing us to gain insights into the activity of the entire NN accelerator.
Our method focuses solely on this duration, which is much shorter than the full inference on the same data. This can be seen in Fig.~\ref{fig:traces} by comparing the Dataflow Loading duration (time for full inference) to the Normalized Size Window (time when all dataflow stages execute simultaneously but on different data).
During an attack, as the exact folding value is unknown, we consistently capture the Normalized Size Window. As a side effect, for the same number of acquired samples, we capture more pattern repetitions compared to the lowest folding configuration.

\subsection{Dataset Preprocessing}\label{exp:preprocess}

We apply preprocessing as described in Section~\ref{methodology}, discarding the dataflow loading phase and retaining only the samples from the Normalized Size Window, followed by normalization.
Next, we fit the PCA using the preprocessed training dataset, employing \textit{scikit-learn} PCA function with default parameters, except for $n_\text{comp}$. After fitting the PCA, we transform (DB$_{\text{training}}$) and (DB$_{\text{test}}$). This PCA transformation reduces the trace dimensionality while amplifying the variance between traces corresponding to different NN parameter configurations.

\begin{figure}
    \centering
    \includegraphics[width=0.9\columnwidth]{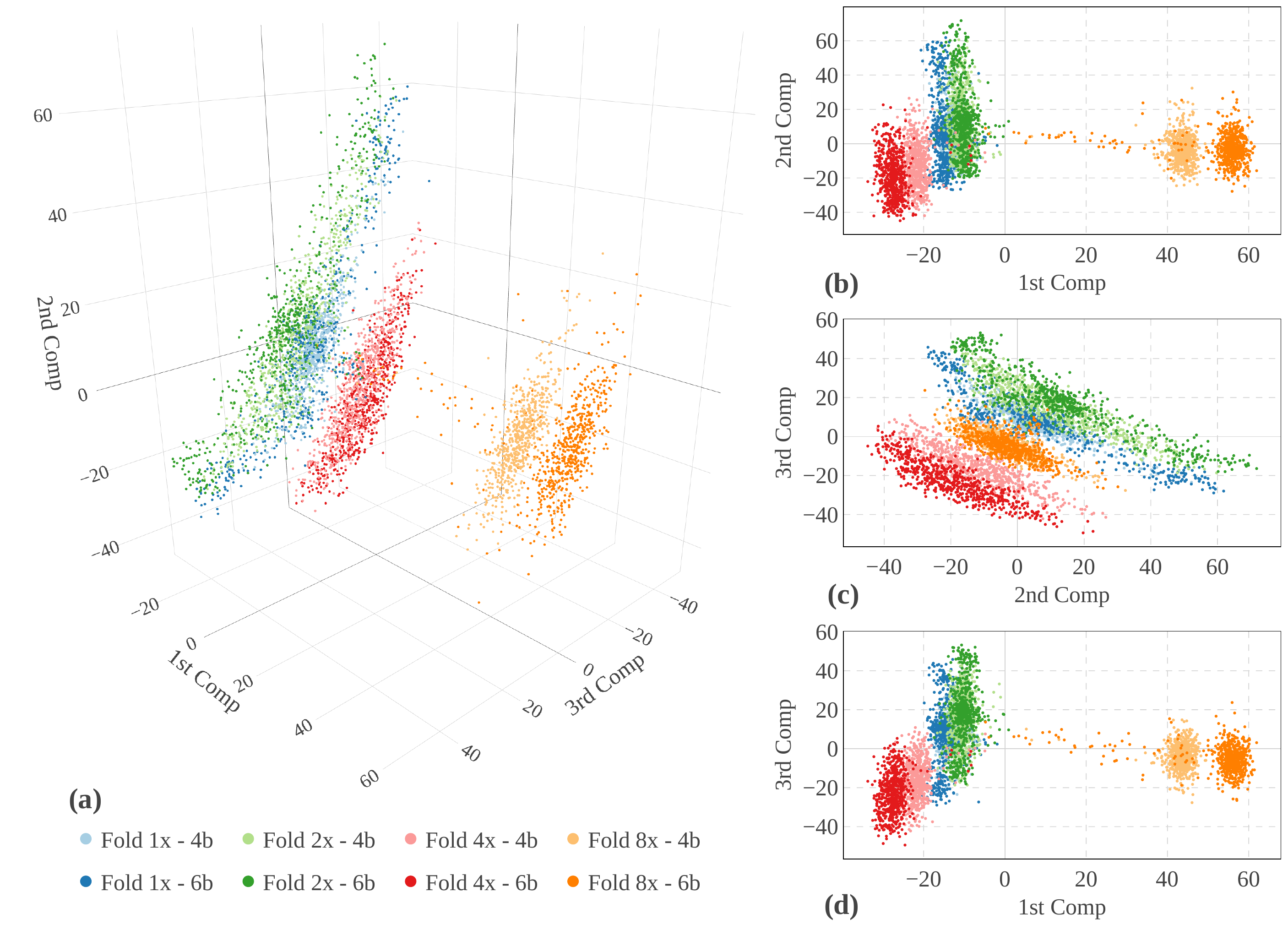}
    \vspace{-0.2cm}
    \caption{First three PCA components. (a) 3D view; projection of components (b) 1 and 2, (c) 2 and 3, (d) 1 and 3. Each point represents a trace.}
    \label{fig:pca}
    \vspace{-0.4cm}
\end{figure}

We plotted the PCA transformation results of the DB$_{\text{training}}$ in Fig.~\ref{fig:pca}. A 3D representation (Fig.~\ref{fig:pca}(a)) is provided alongside 2D projections (Fig.~\ref{fig:pca}(b--d)) for clearer point projections. Each point represents a 129k-sample trace transformed into 3 values, known as components. While more components can be used (see Section~\ref{exp:class}), their graphical representation becomes more complex.
Points are coloured per accelerator configuration, using equivalent light and dark colours for 4-bit and 6-bit quantization, respectively.
In Fig.~\ref{fig:pca}, traces from the same accelerator configuration are clustered together. Dark clusters, separated by folding values, show PCA's ability to detect significant variability among traces for each folding. This distinct clustering will enable the association of a new trace from a blind attack with a cluster through a classifier to recover the corresponding folding.
However, this separation is less pronounced for lower folding values, likely due to greater amplitude variation in traces at higher folding. Nevertheless, these clusters remain separable by a classifier.

To compare with SoA folding extraction~\cite{meyersRevEngNNFolding2022}, we adapt our methodology to use \textit{tsfresh} instead of PCA. We follow a similar method, with feature extraction on the full dataset, and a selection of the top 10 features from our traces, as in~\cite{meyersRevEngNNFolding2022}. 

\subsection{Classifier Exploration}\label{exp:class}

After dimension reduction with PCA, our attack is compatible with any supervised classifier. 
The effectiveness of our method is emphasized by the simplicity of the classifiers we can use. The PCA preprocessing step efficiently transforms the data, separating distinct traces while grouping similar ones. We explore two computationally simple classifiers: k-Nearest neighbors (k-NN) and Random Forest (RF).
Table~\ref{tab:classifer_timings} shows the computation time for each main step of the methodology for both classifiers during the preparation and attack phases, excluding acquisition time. The computations rely on a server with a dual AMD EPYC 7443 24-core CPU at 4 GHz and 512 GB DDR4 3200 MT/s, running CentOS 7.
DB preparation took 15.7 seconds for the k-NN and 17.7 seconds for the RF classifiers when using the PCA, whereas using \textit{tsfresh} took around 4.62 hours, which represents more than 940$\times$ compared to the PCA. The time to run the attack on a single trace was 337ms for k-NN and 421ms for RF using the PCA, where it takes more than 110s to attack one trace with \textit{tsfresh}. This demonstrates that our methodology is less computationally intensive, thanks to the efficiency of PCA dimensionality reduction. This also opens the possibility to improve the results, e.g., using an average factor of 4 with a small cost in time, as also shown in Table~\ref{tab:classifer_timings}. 

\begin{table}[]
    \centering
    \vspace{-0.2cm}
    \caption{Time required per step of our attack methodology.}
    \centering
    \resizebox{0.95\columnwidth}{!}{%
        \begin{tabular}{cc|ccccc|c|}
            \cline{3-8}
             &  & \textbf{DB Build} & \textbf{Norm}. & \textbf{Extract.} & \textbf{Folding} & \textbf{Quant}. & \textbf{Total} \\ \hline \hline
            \multicolumn{1}{|c|}{\multirow{6}{*}{\begin{tabular}[c]{@{}c@{}}\textbf{Preparation} \\\textbf{on DB }[s] \end{tabular}}} & K-NN (tsfresh) & 7.5 & 1.7 & 16650 & 0.85 & 0.95 & 16661 \\ \cline{2-8} 
            \multicolumn{1}{|c|}{} & K-NN (PCA) & 7.5 & 1.7 & 6.5 & 0.85 & 0.85 & 17.4 \\ \cline{2-8}   
            \multicolumn{1}{|c|}{} & K-NN (PCA-A4) & 7.5 & 1.7 & 100 & 0.85 & 0.85 & 110.9 \\ \cline{2-8} 
            \multicolumn{1}{|c|}{} & RF (tsfresh) & 7.5 & 1.7 & 16650 & 0.95 & 0.95 & 16661.1 \\ \cline{2-8}   
            \multicolumn{1}{|c|}{} & RF (PCA) & 7.5 & 1.7 & 8.5 & 1.0 & 1.0 & 19.7 \\ \cline{2-8}
            \multicolumn{1}{|c|}{} & RF (PCA-A4) & 7.5 & 1.7 & 80 & 1.0 & 1.0 & 91.2\\ \hline \hline
            \multicolumn{1}{|c|}{\multirow{6}{*}{\begin{tabular}[c]{@{}c@{}}\textbf{1 trace} \\ \textbf{attack} [s]\end{tabular}}}& K-NN (tsfresh) & - & 0.003 & 110 & 0.25 & 0.25 & 110.5 \\ \cline{2-8}
            \multicolumn{1}{|c|}{} & K-NN (PCA) & - & 0.003 & 0.28 & 0.027 & 0.027 & 0.337 \\ \cline{2-8}
            \multicolumn{1}{|c|}{} & K-NN (PCA-A4) & - & 0.003 & 0.64 & 0.03 & 0.03 & 0.703 \\ \cline{2-8}
            \multicolumn{1}{|c|}{} & RF (tsfresh) & - & 0.003 & 110 & 0.75 & 0.75 & 111.5 \\ \cline{2-8}
            \multicolumn{1}{|c|}{} & RF (PCA) & - & 0.003 & 0.29 & 0.06 & 0.07 & 0.421 \\ \cline{2-8}
            \multicolumn{1}{|c|}{} & RF (PCA-A4) & - & 0.003 & 0.66 & 0.06 & 0.07 & 0.79 \\ \hline
        \end{tabular}%
        }
    \label{tab:classifer_timings}
    \vspace{-0.4cm}
\end{table}

In the following, we continue to apply our methodology, as presented in Section IV. Both classifiers will be fed with DB$_{\text{training}}$ for tuning and validated with DB$_{\text{test}}$, through a grid search process. 
As mentioned in Section~\ref{methodology}, we use multiple lightweight classifiers to recover parameters. One classifier recovers the folding configuration, while another handles quantization. This approach offers better accuracy than a single, larger classifier trying to distinguish all classes simultaneously.

\subsubsection{k-nearest neighbors (k-NN)}

The first classification approach employs k-NN classifiers, which are deterministic and computationally lightweight. Classification requires a preprocessed new trace, a labeled DB for comparison, and a $k$ number of neighbors for similarity-based classification. Only $k$ and $n_\text{comp}$ need to be explored during the grid search.
We limit grid search exploration to $n_\text{comp\_max} = 50$ and explore $k$ from 3 to 19.
Based on the grid search results, we set the attack parameters to $n_\text{comp}$ = 12 and $k$ = 13.

We evaluate the final accuracy of our attack using DB$_{\text{test}}$, first applying the PCA and then using the classifiers. The results are shown in Table~\ref{tab:classif-detail}, which illustrates the classification accuracy for folding and quantization across different values of $n_{\text{average}}$.
For folding, our classifier achieves 91.84\% accuracy without averaging and 96.78\% accuracy with averaging. This demonstrates the effectiveness of PCA and shows that the NN accelerator leaks sufficient information, even with a limited number of samples relative to the Normalized Window Size and a simple k-NN classifier. Importantly, our method is data-independent, as evidenced by the impact of averaging independent traces, i.e., traces obtained using different inputs.

For better understanding, Table~\ref{tab:classif-detail} shows how the classifier determines the correct implementation parameters from a trace. For instance, when Folding 1$\times$ is combined with 4 bits of quantization, the k-NN dedicated to quantization outputs the correct result 65.9\% of the time. This indicates that classification is more challenging for quantization with Folding 1$\times$ and 2$\times$.
This can be explained by the fact that the clusters for these folding values are less distinct in the PCA space, as mentioned in Section~\ref{exp:preprocess} and visible in Fig~\ref{fig:pca}.
However, with the PCA, the recovery is globally better than using \textit{tsfresh} with 90.91\%, both having some difficulties in some cases. To improve these results, a stronger classifier could be used such as RF.

\subsubsection{Random Forest}

Unlike k-NN, which may struggle with closely clustered data points, RF should be more effective by leveraging an ensemble of decision trees to capture complex decision boundaries, enhancing prediction accuracy, even when the data clusters are not visually separable.
We limit grid search exploration with $n_\text{comp\_max} = 50$. The explored parameters are
$n_\text{estimators}$ in $\{100, 200, 400\}$, and $min_\text{samples\_split}$ in $\{2, 5, 10\}$.
The best accuracy is found for the parameters: $n_\text{comp}=40$, $n_\text{estimators}=400$ and $min_\text{samples\_split}=5$.
Table~\ref{tab:classif-detail} shows significant improvements over k-NN, achieving 99.7\% accuracy for folding and over 93.7\% for quantization without any averaging, and 99.96\% for both elements when using an averaging of 4 traces (PCA-A4). 
Regarding the detail of parameter recovery with respect to the implemented DNN, Table~\ref{tab:classif-detail} highlights that even for Folding 1$\times$ and 4-bit quantization, the classification accuracy exceeds 96\%, confirming that RF can efficiently identify closely clustered data points while maintaining limited computation time.

\begin{table}[]
\centering
    \vspace{-0.2cm}
\caption{Classifier accuracy vs. HW parameters and label.}
    \centering
    \resizebox{0.95\columnwidth}{!}{%
    \begin{tabular}{|c|c|cccc|cccc|c|}
    \hline
    \multirow{2}{*}{\begin{tabular}[c]{@{}c@{}}\textbf{Accelerator}\\ \textbf{parameters}\end{tabular}} & Quant. & \multicolumn{4}{c|}{4 bits} & \multicolumn{4}{c|}{6 bits} & \multirow{2}{*}{\textbf{Avg}} \\ \cline{2-10} 
     & Fold. & 1x & 2x & 4x & 8x & 1x & 2x & 4x & 8x & \\ \hline\hline
    \multirow{2}{*}{\begin{tabular}[c]{@{}c@{}}\textbf{K-NN (tsfresh)}\\ \textbf{accuracy} {[}\%{]}\end{tabular}} & Quant. & 100 & 84.1 & 99.3 &  \textbf{100} & 94.1 & 99.2 & 88.3 & 97.6 & \multirow{2}{*}{90.91} \\ \cline{2-10} 
     & Fold. & 91.6 & 70.2 & 96.7 & 99.4 & 80 & 96.7 & 75.4 & 81.9 & \\ \hline
     \multirow{2}{*}{\begin{tabular}[c]{@{}c@{}}\textbf{K-NN (PCA)}\\ \textbf{accuracy} {[}\%{]}\end{tabular}} & Quant. & 65.9 & 85.7 & 89.3 & 98.7 & 71.9 & 77.8 & 99.3 & 96.9 & \multirow{2}{*}{91.84} \\ \cline{2-10} 
     & Fold. & 93.5 & 98.7 & \textbf{100} & \textbf{100} & 99.3 & 92.4 & \textbf{100} & \textbf{100} & \\ \hline
     \multirow{2}{*}{\begin{tabular}[c]{@{}c@{}}\textbf{K-NN (PCA-A4)}\\ \textbf{accuracy} {[}\%{]}\end{tabular}} & Quant. & 84.6 & 92.4 & 100 & 100 & 91.7 & 79.8 & 100 & 100 & \multirow{2}{*}{96.78} \\ \cline{2-10} 
     & Fold. & \textbf{100} & \textbf{100} & \textbf{100} & \textbf{100} & \textbf{100} & \textbf{100} & \textbf{100} & \textbf{100} & \\ \hline\hline
    \multirow{2}{*}{\begin{tabular}[c]{@{}c@{}}\textbf{RF (tsfresh)}\\ \textbf{accuracy} {[}\%{]}\end{tabular}} & Quant. & \textbf{100} & 90 & \textbf{100} & 99.4 & 96.4 & \textbf{100} & 97 & 98.8 & \multirow{2}{*}{95.72} \\ \cline{2-10} 
     & Fold. & 98 & 84.7 & 94.5 & 98.1 & 92.3 & 99.2 & 86.1 & 97 & \\ \hline
     \multirow{2}{*}{\begin{tabular}[c]{@{}c@{}}\textbf{RF (PCA)}\\ \textbf{accuracy} {[}\%{]}\end{tabular}} & Quant. & 82.7 & 90.3 & 94.6 & 96.3 & 91.5 & 98.6 & 98.1 & 97.5 & \multirow{2}{*}{96.70} \\ \cline{2-10} 
     & Fold. & \textbf{100} & 98.2 & \textbf{100} & \textbf{100} & \textbf{100} & 99.4 & \textbf{100} & \textbf{100} & \\ \hline
     \multirow{2}{*}{\begin{tabular}[c]{@{}c@{}}\textbf{RF (PCA-A4)}\\ \textbf{accuracy} {[}\%{]}\end{tabular}} & Quant. & \textbf{100} & \textbf{99.4} & \textbf{100} & \textbf{100} & \textbf{100} & \textbf{100} & \textbf{100} & \textbf{100} & \multirow{2}{*}{\textbf{99.96}} \\ \cline{2-10} 
     & Fold. & \textbf{100} & \textbf{100} & \textbf{100} & \textbf{100} & \textbf{100} & \textbf{100} & \textbf{100} & \textbf{100} & \\ \hline
    \end{tabular}%
    }
 \vspace{-0.4cm}
\label{tab:classif-detail}
\end{table}

\section{Conclusion} \label{conclusion}
This paper proposes an SCA attack methodology for FPGA-based DNN accelerators using remote sensors. Our approach leverages PCA dimensionality reduction and demonstrates its efficiency in enabling the use of lightweight classifiers for parameter recovery.
Our attack model is more robust and coherent in a remote context compared to SoA solutions and is data-independent. Experiments show the attack phase requires only 400 ms to recover over 95\% of folding and quantization parameters of a FINN-based CNN accelerator using an RF classifier. This is achieved by acquiring one trace with a duration equal to the slowest dataflow stage, validating our methodology. Compared to SoA attacks based on \textit{tsfresh}, our method requires 940$\times$ and 110$\times$ less time, for preparation and attack phases, respectively, and gives better results even without averaging traces. By leveraging this time gain, we could use an average of 4 traces to fully recover all the HW configurations tested within only 800 ms for the attack phase.
Lastly, the core of our method relies on PCA, an unsupervised process that allows for the use of unsupervised classifiers to distinguish between different implementations, though recovering unknown parameters may require additional effort.

\section*{Acknowledgments}
This work is partially funded by the French \emph{Agence Nationale de la Recherche} (ANR) Young Researchers (JCJC) program, under grant number ANR-21-CE39-0018 (project ATTILA).

\bibliographystyle{IEEEtran}
\bibliography{sources}
\end{document}